"TRANSITION FROM POISSON TO GAUSSIAN UNITARY STATISTICS :
THE TWO-POINT CORRELATION FUNCTION"

Hervé KUNZ
Ecole Polytechnique Fédérale de Lausanne
Institut de Physique Théorique
PHB-Ecublens
CH-1015 LAUSANNE, Switzerland
E-mail : Herve.Kunz@epfl.ch

and

Boris SHAPIRO
Technion - Israel Institute of Technology
Department of Physics
HAIFA 32000, Israel
E-mail : Shapiro Boris <boris@physics.technion.ac.il>

1. INTRODUCTION

Energy spectra of complex quantum systems often exhibit a crossover from the Poisson statistics of uncorrelated levels to the Wigner-Dyson statistics, with its characteristic level repulsion and rigidity. There is therefore some interest in designing and studying random matrix ensembles which could interpolate between these two extremes [1]. One such generalisation goes back to the work of Rosenzweig and Porter [2] on complex atomic spectra. They have studied, numerically, random matrices with enhanced (as compared with the standard ensembles) weight for the diagonal elements. A slight generalisation of the Rosenzweig-Porter matrix model leads to the following ensemble for N x N hermitian matrices :

$$H = A + \frac{\lambda}{N^{\alpha}} V \qquad (1.1)$$

where A is a random diagonal matrix, i.e.

$$A = \text{diag}\left(a_1 L\ a_N\right) \qquad (1.2)$$

with some probability distribution p(a) for its statistically independent elements. V is a real or complex hermitian matrix, whose matrix elements are supposed to be independent, with a gaussian distribution of unit variance. There is no loss of generality, by considering A to be diagonal as long as we are interested only by correlations functions of the eigenvalues of H.

Different behaviour can be expected by varying the exponent $\alpha$. If $\alpha = \frac{1}{2}$, one expects GOE or GUE statistics whereas if $\alpha > 1$ the statistics should be of the Poisson type. It seems that the only way to obtain new statistics is to choose $\alpha = 1$. For technical reasons, up to now only the case of a complex hermitian V matrix has been treated. In an important work, which has inspired us, Brezin and Hikami [3], abbreviated in the following BH, showed that when $\alpha = \frac{1}{2}$, and for any fixed matrix A, the statistics is GUE. Pandey [4] has discussed the case $\alpha = 1$ in great generality, using a Fokker-Planck equation for the probability distribution of eigenvalues. In the complex case, he could successfully use the fact that this equation is integrable and gave an expression for the two-point correlation function and the spectral form factor in a number of cases, including the one discussed here.

Unfortunately, no derivation of the result was given. More recently, Guhr and Guhr and Müller-Groeling [5] have derived a double-integral representation of the two-point correlation function, using a variant of the supersymmetric technique. In our opinion, however, this work rests on unproven prescription for obtaining the correlation functions from averages of advanced Green's functions only.

Finally, Altland et al [6] using perturbation theory in $\lambda$ obtained a closed expression for the two-point correlation function. However, since they kept the imaginary part of the energy non zero, before going to the large N limit their expression gives only an "overview" of the picture : it is limited to large distances and completely misses the fine structure of the correlation function at shorter distances.

In the present paper, we look again at this problem (in the complex case), using BH strategy. We show first that the density of states $\nu(E)$ is given by p, and that if we unfold the two-point correlation function, in the usual way in energy, and also in the coupling constant $\lambda$ replacing it by $\dfrac{\lambda}{\nu(E)}$, then we get a universal function, which coincides with that given by Pandey. Our approach gives more naturally the spectral form factor. Both quantities can be expressed as integrals of Bessel functions with imaginary arguments. Quite generally, we find that the model gives repulsion of levels at short distances and attraction of levels at long distances. The small coupling limit (close to Poisson) and the strong coupling limit (close to GUE) are described explicitly. In the small coupling limit, we recover the results of Leyvraz and Seligman [7] obtained by a perturbation technique.

## 2. PROBABILITY DISTRIBUTION OF EIGENVALUES AND DENSITY OF STATES

We consider an Hamiltonian H of the form

$$H = A + V \tag{2.1}$$

A being a fixed matrix and V a matrix belonging to the unitary ensemble of probability distribution

$$P(V) \sim \exp\left[-\frac{1}{2\sigma} \operatorname{tr} V^2\right] \tag{2.2}$$

$\sigma$ being a parameter whose dependence on N will be fixed later.

The probability distribution of H will be therefore

$$P(H) \sim \exp\left[-\frac{1}{2\sigma} \operatorname{tr} H^2 + \frac{1}{\sigma} \operatorname{tr} AH\right] \tag{2.3}$$

In order to obtain the probability distribution $W(\lambda_1...\lambda_N)$ for the eigenvalues $\lambda_j$ of H, one integrates over the unitary matrices U diagonalising H, by using the Harish-Chandra, Itzykson-Zuber identity [1]

$$\int dU \, \exp\left[\frac{1}{\sigma} \operatorname{tr} AU\lambda U^+\right] = \frac{c}{\Delta(\lambda)\,\Delta(a)} \det \exp\left(\frac{a_i \, \lambda_j}{\sigma}\right) \tag{2.4}$$

the variables $a_j$ being the eigenvalues of the matrices A and $\Delta(a)$ being the van der Monde determinant

$$\Delta(a) = \prod_{i<j}^{N} \left(a_i - a_j\right) \tag{2.5}$$

c is a constant depending on N.

If we use the identity

$$\int d\lambda_1 \ldots d\lambda_N \, \Delta(\lambda) \exp\left[-\frac{1}{2\sigma} \sum_{i=1}^{N} (\lambda_i - a_i)^2\right] = \Delta(a)\,(2\pi\sigma)^{\frac{N}{2}} \tag{2.6}$$

whose proof is given in the appendix, for completeness, we get for the normalised probability distribution W of the eigenvalues

$$W(\lambda_1...\lambda_N) = \frac{1}{N!} \left(\frac{1}{2\pi\sigma}\right)^{\frac{N}{2}} \frac{\Delta(\lambda)}{\Delta(a)} \exp\left[-\frac{1}{2\sigma} \sum_{i=1}^{N} \left(a_i^2 + \lambda_i^2\right)\right] \det \exp\left(\frac{1}{\sigma} a_i \lambda_j\right) \quad (2.7)$$

In what follows, we will always consider average values of symmetric functions of the eigenvalues $F(\lambda_1, ..., \lambda_N)$, with respect to the distribution W. In such a case the average value can be simply written as

$$\langle F \rangle_W = \left(\frac{1}{2\pi\sigma}\right)^{\frac{N}{2}} \frac{1}{\Delta(a)} \int d\lambda_1...d\lambda_N \ F(\lambda_1...\lambda_N) \ \Delta(\lambda) \ \exp\left[-\frac{1}{2\sigma} \sum_i (\lambda_i - a_i)^2\right] \quad (2.8)$$

because the determinant of the matrix $\exp\left(\frac{1}{\sigma} a_i \lambda_j\right)$ can be simply replaced by

$$N! \exp\left[\frac{1}{\sigma} \sum_i a_i \lambda_i\right].$$

Thus the full average of a symmetric function F over the ensemble of H matrices defined by Eq. (2.1) will be

$$\overline{F} = \int da_1 L \ da_N \prod_{i=1}^{N} p(a_i) \ \langle F \rangle_W$$

since we consider a model where all the $a_j$ are statistically independent and identically distributed.

Following BH, it appears more convenient to compute first the Fourier transforms of correlation functions, i.e.

$$C_1(t) = \sum_k \exp\left(it \ \lambda_k\right)$$

$$C_2(t, t') = \sum_{k \neq l} \exp\left(it \ \lambda_k + it' \ \lambda_l\right)$$

(2.9)

Since from (2.6) follows that

$$\int d\lambda_1 \, L \, d\lambda_N \, \exp\left[\sum_i b_i \lambda_i\right] \Delta(\lambda) \exp\left[-\frac{1}{2\sigma} \sum_i (\lambda_i - a_i)^2\right]$$

(2.10)

$$= \Delta(a + b\sigma)(2\pi\sigma)^{\frac{N}{2}} \exp\left[\sum_i a_i b_i + \frac{\sigma}{2} \sum_i b_i^2\right]$$

we get

$$\langle C_1 \rangle_W = \exp-\frac{\sigma t^2}{2} \sum_k \exp(it a_k) \prod_{j \neq k} \left(1 + \frac{it\sigma}{a_k - a_j}\right)$$

and (2.11)

$$\langle C_2 \rangle_W = \exp-\frac{\sigma}{2}(t^2 + t'^2) \sum_{kl} \exp(it a_k + it' a_l) \prod_{j \neq k}\left(1 + \frac{it\sigma}{a_k - a_j}\right) \prod_{j \neq l}\left(1 + \frac{it'\sigma}{a_l - a_j}\right)$$

$$\times F(a_k, a_l) \quad (2.12)$$

where $F(a_k, a_l) = \dfrac{a_l - a_k}{a_l - a_k - it\sigma} \dfrac{a_l - a_k + it'\sigma - it\sigma}{a_l - a_k + it'\sigma}$ (2.13)

It remains to average over the a's. For this purpose we follow BH and use the following integral representations for $\langle C_1 \rangle_W$ and $\langle C_2 \rangle_W$.

$$\langle C_1 \rangle_W = \exp\left(-\frac{\sigma t^2}{2}\right) \frac{1}{it\sigma} \oint_{\Gamma_R} \frac{dz}{2\pi i} \exp(itz) \prod_{j=k}^N \left(1 + \frac{it\sigma}{z - a_j}\right) \quad (2.14)$$

and

$$\langle C_2 \rangle_W = -\exp\left[-\frac{\sigma}{2}(t^2 + t'^2)\right] \frac{1}{tt'\sigma^2} \oint_{\Gamma_R} \frac{dz}{2\pi i} \oint_{\Gamma_R} \frac{dz'}{2\pi i} \exp[i(tz + t'z')]$$

$$\times \prod_{j=1}^N \left(1 + \frac{it\sigma}{z - a_j}\right)\left(1 + \frac{it'\sigma}{z - a_j}\right) F(z, z') \quad (2.15)$$

$\Gamma_R$ being the contour formed by a rectangle of vertical size $2\epsilon$ and horizontal one $2R$ encircling the origin. One needs to choose $R > \sup_j |a_j|$, and in (2.15) one needs to impose that $|t| > 2\epsilon$ and $|t'| > 2\epsilon$, in order to avoid the poles in the function $F(z,z')$ defined by (2.13)

$$F(z, z') = \frac{z'-z - i\sigma(t-t')}{z'-z + it'\sigma} \cdot \frac{z'-z}{z'-z - it\sigma} \tag{2.16}$$

In this way, one gets for the Fourier transform of the density of states

$$\overline{C}_1(t) = \exp-\frac{\sigma}{2} t^2 \frac{1}{i\sigma t} \oint_{\Gamma_R} \frac{dz}{2\pi i} \exp(itz) \left(1 + i\sigma t < \frac{1}{z-a} >\right)^N \tag{2.17}$$

where from now, $<\cdot>$ will designate an average with respect to the variable a, i.e.

$$< \frac{1}{z-a} > = \int da\ p(a)\ \frac{1}{z-a} \tag{2.18}$$

Up to now, the validity of Eq. (2.17) is guaranteed only for a probability distributions of the a's of bounded support. However, since, as it is easily checked, we can take the limit $R \to \infty$ in this formula, the contour $\Gamma_R$ becoming the contour $\Gamma$ made of the two lines $x \pm i\varepsilon$. In this way the formula becomes valid for distribution $p(a)$ of unbounded support.

Equation (2.17) holds for any N. It is instructive to consider the $N \to \infty$ limit, in order to exhibit the importance of the parameter $\sigma$.

We can rewrite Eq. (2.17) as

$$\frac{1}{N} \overline{C}_1(t) = \frac{1}{N} \exp\left[-\frac{\sigma t^2}{2}\right] \oint_\Gamma \frac{dz}{2\pi i} e^{itz} \sum_{j=0}^{N} \binom{N}{j} (i\sigma t)^{j-1} < \frac{1}{z-a} >^j \tag{2.19}$$

The $j = 0$ term vanishes. The $j = 1$ term does not depend on N. The behaviour of higher terms depends on the value of the parameter $N\sigma$. Since we are interested in the case where $N\sigma$ tends to zero, we see that all terms beyond $j \geq 2$ disappear so that

$$\lim_{N \to \infty} \frac{1}{N} \overline{C}_1(t) = \oint_\Gamma \frac{dz\ e^{itz}}{2\pi i} < \frac{1}{z-a} > = <e^{ita}> \tag{2.20}$$

Hence the average density of states $\nu(\lambda)$ is simply given by $p(\lambda)$ [6] in our case.

## 3. THE SPECTRAL FORM FACTOR

We want now to compute the $N = \infty$ limit of the correlation function $C(t, t')$ defined by

$$NC(t, t') = \overline{C}_2(t, t') - \overline{C}_1(t)\overline{C}_1(t') \tag{3.1}$$

From Eq. (2.14) and (2.15), we see that we can represent it in the form

$$C(t, t') = [K_1(t, t') + K_2(t, t')] \exp\left[-\frac{\sigma}{2}\left(t^2 + t'^2\right)\right] \tag{3.2}$$

with

$$K_1(t, t') = \frac{1}{N\tau\tau'} \oint_\Gamma \frac{dz}{2\pi i} \oint_\Gamma \frac{dz'}{2\pi i} \exp\left[i(tz + t'z')\right] \left[g^N(z, z') - \rho^N(z/\tau)\rho^N(z'/\tau')\right] \tag{3.3}$$

and

$$-K_2(t, t') = \frac{1}{N(\tau + \tau')} \oint_\Gamma \frac{dz}{2\pi i} \oint_\Gamma \frac{dz'}{2\pi i} \exp\left[i(tz + t'z')\right] g^N(z, z')$$

$$\left[\frac{1}{z - z' + \tau} + \frac{1}{z' - z + \tau'}\right] \tag{3.4}$$

where we used the notation

$$\tau = it\sigma$$
$$\tau' = it'\sigma \tag{3.5}$$

$$g(z, z') = 1 + \tau \alpha(z) + \tau' \alpha(z') + \frac{\tau\tau'}{z - z'}[\alpha(z') - \alpha(z)] \tag{3.6}$$

with $\quad \alpha(z) = \left\langle \frac{1}{z - a} \right\rangle \tag{3.7}$

and
$$\rho(z/\tau) = 1 + \tau \alpha(z)$$
$$\rho(z'/\tau') = 1 + \tau' \alpha(z') \tag{3.8}$$

From now on, we will take $\sigma = \dfrac{\lambda^2}{N^2}$ (3.9)

In order to get a finite limit for C, it is appropriate to choose the scale, defined by

$$t = \frac{T}{2} + Ns$$
$$t' = \frac{T}{2} - Ns$$
(3.10)

and keep T and s fixed when N tends to $\infty$. We then choose $\varepsilon = \dfrac{\delta}{N}$ in the contour $\Gamma$ and represent the variables z and z' by

$$z = x + \frac{y}{2N} - \frac{i\delta}{N} q$$
$$z' = x - \frac{y}{2N} - \frac{i\delta}{N} q'$$
(3.11)

q and q' being equal to $\pm 1$, depending on the branch of $\Gamma$ on which the variables stay. We note that the function $\alpha(z)$ defined by Eq. (3.7) is such that

$$\alpha\left(x - \frac{i\delta q}{N}\right) = \gamma_q(x) + 0\left(\frac{1}{N}\right)$$

with (3.12)

$$\gamma_q(x) = P \int \frac{p(a)da}{x-a} + i\pi q\, p(x)$$

Since $\tau = \dfrac{is\lambda^2}{N}$ and $\tau' = \dfrac{-is\lambda^2}{N}$, at the order $\dfrac{1}{N^2}$, it follows that

$$\rho(z/\tau) = 1 + \frac{is\lambda^2}{N} \gamma_q(x) + 0\left(\frac{1}{N}\right)$$
$$\rho(z'/\tau') = 1 - \frac{is\lambda^2}{N} \gamma_{q'}(x) + 0\left(\frac{1}{N}\right)$$
(3.13)

and if q = q'

$$g(z, z') = 1 + O\left(\frac{1}{N^2}\right) \qquad (3.14)$$

but if $q = -q'$

$$g(z, z') = 1 - \frac{2\pi s \lambda^2}{N} p(x)\left[1 + \frac{s\lambda^2 iq}{qy - 2i\delta}\right] + O\left(\frac{1}{N^2}\right) \qquad (3.15)$$

From these results, one can derive the desired asymptotic behaviour

$$K_{\frac{1}{2}}(t, t') = \sum_q \int dx\, p(x)\, e^{iTx - 2\pi\lambda^2 sqp(x)} L_{\frac{1}{2}} \qquad (3.16)$$

where

$$L_{\frac{1}{2}} = \int_{-\infty}^{+\infty} \frac{dy}{2\pi}\, e^{ib(y-i\delta)} \left[\exp\left(-\frac{ia}{y-i\delta}\right) - 1\right] F_{\frac{1}{2}} \qquad (3.17)$$

with $F_1 = \dfrac{1}{a}$

$$F_2 = \frac{1}{4\pi p} \frac{1}{\left[y + i\left(\frac{\lambda^2 sq}{2} - \delta\right)\right]^2} \qquad (3.18)$$

and
$$\begin{aligned} a &= \pi\lambda^4 s^2 p \\ b &= 2sq \end{aligned} \qquad (3.19)$$

It remains to evaluate these integrals. One first notice that they vanish when b is negative. This implies that in Eq. (3.16), one can drop the summation on q and replace sq by $|s|$.

$L_1$ can be evaluated in closed form and is given by

$$L_1 = \frac{1}{\sqrt{ab}}\, I_1\!\left(2\sqrt{ab}\right) \qquad (3.20)$$

$I_1(z)$ being the modified Bessel function given by the series

$$I_1(z) = \sum_{j=0}^{\infty} \frac{\left(\frac{z}{2}\right)^{2j+1}}{j!(j+1)!} \tag{3.21}$$

In evaluating $L_2$ for b positive, one notices that when b is positive $\frac{\lambda^2 sq}{2} - \delta = \frac{\lambda^2 |s|}{2} - \delta$ is positive because the formula for $K_2$ is valid under this constraint only, for N very large because of the condition $|t| > \frac{2\delta}{N}$ and $|t'| \geq \frac{2\delta}{N}$ that we needed to impose in Eq. (2.15). One can then express $L_2$ in the form:

$$L_2 = -\frac{b^2}{4\pi p} \sqrt{\frac{a}{b}} \int_0^\infty \frac{dt\, t}{\sqrt{t+1}} I_1\left(2\sqrt{ab(t+1)}\right) \exp\left(-t\,\frac{b^2\lambda^2}{4}\right) \tag{3.22}$$

Grouping all these results we can finally write the correlation function C in the form

$$C(t, t') = -\int dx\, p(x)\, e^{iTx}\, S\left(u = \frac{|s|}{p(x)} / \Lambda = \lambda p(x)\right) \tag{3.23}$$

where the function $S(u/\Lambda)$ is given by

$$S(u/\Lambda) = -\frac{2}{\gamma} I_1(\gamma) \exp\left[-2\pi\Lambda^2 u - \Lambda^2 u^2\right]$$
$$+ \frac{u}{2\pi} \gamma \int_1^\infty dt\, (t^2-1)\, I_1(\gamma t)\, \exp\left[-t^2\Lambda^2 u^2 - 2\pi\Lambda^2 u\right] \tag{3.24}$$

with $\quad \gamma = \sqrt{8\pi\Lambda^4 u^3} \tag{3.25}$

These expressions are our main result. We prove now that **S is the unfolded spectral form factor.**

Indeed, by definition, C(t, t') is related to the eigenvalues correlation functions.

$$\rho_2(x,x') = \sum_{k \neq 1} \delta(x-\lambda_k) \, \delta(x'-\lambda_1)$$

$$\rho_1(x) = \sum_k \delta(x-\lambda_k)$$

(3.26)

by the equation

$$C(t, t') = \int dx \int dx' \, e^{i(tx + t'x')} \, \frac{1}{N} \left[ \rho_2(x,x') - \rho_1(x) \, \rho_1(x') \right] \qquad (3.27)$$

Let us make in this integral the change of variables ($v(E)$ being the density of states)

$$x = E + \frac{r}{2N v(E)}$$

$$x' = E - \frac{r}{2N v(E)}$$

(3.28)

and go from the variables $(t, t')$ to $(T, s)$ as defined by equation (3.9).

Then we can see that we can express C as

$$C(t,t') = -\int dE \, v(E) \, e^{iTE} \int dr \, Y_N(E,r) \, \exp\left( ir \frac{s}{v(E)} \right) \qquad (3.29)$$

where

$$Y_N(E,r) = \frac{1}{[N v(E)]^2} \left\{ \rho_1\left(E + \frac{r}{2N v(E)}\right) \rho_1\left(E - \frac{r}{2N v(E)}\right) \right.$$

$$\left. -\rho_2\left(E + \frac{r}{2N v(E)}, E - \frac{r}{2N v(E)}\right) \right\} \qquad (3.30)$$

is the unfolded (in energy) connected two-point correlation function.

Notice that in our case $v(E) = p(E)$.

Our result for the asymptotic behaviour for C can therefore be interpreted as saying that if we unfold $Y_N(E, r)$ in the coupling constant too, i.e. replace $\lambda$ by $\Lambda \setminus v(E)$, then $Y_N(E, r)$ tends to a function $Y(r/\Lambda)$ independent of $v$.

and

$$\int dr \; Y(r/\Lambda) \exp(iru) = S(u/\Lambda) \tag{3.31}$$

is the spectral form factor. In fact, we found $S(u) = S(|u|)$. We will now analyse in more detail the unfolded form factor $S(u/\Lambda)$ and the connected correlation function $Y(r/\Lambda)$.

## 4. TWO-POINT CORRELATION FUNCTION. STRONG AND WEAK COUPLING LIMITS

Let us first give some properties of S and Y valid for any coupling $\Lambda$.

Developing in series $I_1(\gamma t)$ in Eq. (3.24), and integrating over t, one can rewrite S in the form

$$S = \frac{1}{2\pi\Lambda^2 u} \left[ \sum_{k=2}^{\infty} k \left(\frac{2\pi}{u}\right)^{\frac{k}{2}} I_k(\gamma) \right] \exp\left[-\Lambda^2 u^2 - 2\pi\Lambda^2 u\right] \qquad (4.1)$$

where $I_k$ is the modified Bessel function of order k.

This expression shows that S is positive, a property possessed by the GUE form-factor. Since the correlation function Y is given by

$$Y(r) = \frac{1}{\pi} \int_0^\infty \cos ur \, S(u) \, du \qquad (4.2)$$

we see immediately that $Y(o) > 0$ and $Y''(o) < 0$, which show that we have **level repulsion at short distances.**

Consider now the long distance behaviour. From (4.1) one can see that for small u

$$S(u) \sim 2\pi\Lambda^2 u \qquad (4.3)$$

This implies that $\int dr\, Y(r) = 0 \qquad (4.4)$

to be contrasted with the sum rule

$$\int dr\, Y(r) = 1 \qquad (4.5)$$

in the GUE case, characteristic of the long range coulomb repulsion of the levels.

Moreover, an integration by parts in (4.2) gives

$$Y(r) = -\frac{2\Lambda^2}{r^2} - \frac{1}{\pi r^2} \int_0^\infty \cos ur \, S''(u) \, du \qquad (4.6)$$

This shows that we have an **attraction of levels at long distances**, and that correlations decay very slowly, like $\frac{1}{r^2}$.

In order to study the asymptotic behaviour of S and Y, it is helpful to look for another integral representation of these quantities.

It can be easily checked from the series expansion of the Bessel function $I_k$, that we have

$$\left(\frac{2\pi}{u}\right)^{\frac{k}{2}} I_k(\gamma) = \oint_{|z|>1} \frac{dz}{2\pi i} \frac{1}{z^{k+1}} \exp\left[\frac{\Lambda^2 u^2}{z} + 2\pi\Lambda^2 uz\right] \qquad (4.7)$$

Inserting this representation in the equation (4.1) for S, we can write it in the form

$$S = \oint_{|z|>1} \frac{dz}{2\pi i} \frac{1}{z} \frac{1}{z-1} \left(1 - \frac{u}{2\pi z^2}\right) \exp\left[-\Lambda^2 u^2 \left(1-z^{-1}\right) - 2\pi\Lambda^2 u(1-z)\right] \qquad (4.8)$$

when $\frac{u}{2\pi}>1$, we choose for the integration path on z a circle of radius $\sqrt{\frac{u}{2\pi}}$. We can do the same choice when $\frac{u}{2\pi}<1$, if we take into account the contribution of the pole at z = 1. In this way, one gets

$$S(u/\Lambda)-S(u/\infty) = -\frac{2}{\pi} \int_{-1}^{1} dy \, \frac{\left[2y\sqrt{\frac{u}{2\pi}} + 1\right]\sqrt{1-y^2}}{\frac{u}{2\pi} + 2y\sqrt{\frac{u}{2\pi}} + 1} \exp\left[-2\pi\Lambda^2 u \left[\frac{u}{2\pi} + 2y\sqrt{\frac{u}{2\pi}} + 1\right]\right]$$

$$(4.9)$$

where

$$S(u/\infty) = \theta\left(1-\frac{u}{2\pi}\right)\left(1-\frac{u}{2\pi}\right) \qquad (4.10)$$

is the GUE form factor.

The corresponding equation for the correlation function Y is

$$Y(r/\Lambda) - Y(r/\infty) = -\frac{4}{\pi} \int_0^\infty dx \cos 2\pi rx \int_{-1}^1 \frac{(2y\sqrt{x}+1)(1-y^2)^{\frac{1}{2}}}{x+2y\sqrt{x}+1}$$

$$\exp\left[-(2\pi\Lambda)^2 \times \left[x+2\sqrt{x}\ y+1\right]\right] \quad (4.11)$$

and

$$Y(r/\infty) = \left(\frac{\sin \pi r}{\pi r}\right)^2 \quad (4.12)$$

This expression coincides with that of Pandey [4] if we identify our $\Lambda^2$ with half his.

This representation of Y is useful to analyse the **large $\Lambda$ limit**, because $\Lambda^2$ appears as the parameter of a Laplace transform. The dominant contributions to the integral will come therefore from the neighbourhood of $x = 0$ and of $x = 1$ and $y = -1$.

A careful analysis shows that the contribution from $x = 0$ gives :

$$\delta Y_1 = -2 \frac{\Lambda^2}{(2\pi\Lambda^2)^2 + r^2} + 0\left(\Lambda^{-3}\right)$$

and the contribution from $x = 1$, $y = -1$ gives

$$\delta Y_2 = +2 \frac{2}{\Lambda^2 \pi^3} \operatorname{Re} \exp(2\pi ir) \int_0^\infty dy \int_{-\infty}^{+\infty} dx \exp\left(\frac{y^2}{x^2+y^2}\right)$$

$$\exp\left[-\left(x^2+y^2\right) + i\frac{4\pi u}{\sqrt{2}}\frac{r}{\Lambda}\right]$$

$$+ 0\left(\Lambda^{-5/2}\right)$$

Evaluating the integral and adding these contributions we find that

$$Y(r) = -\frac{2\Lambda^2}{\left(2\pi\Lambda^2\right)^2 + r^2} + \frac{1}{2(\pi r)^2}\left[\cos 2\pi r \exp\left[-\left(\frac{r}{\Lambda}\right)^2\right] - 1\right]$$

$$+ 0\left(\Lambda^{-5/2}\right) \tag{4.13}$$

It is important to note that this asymptotic expansion holds uniformly in r, i.e. the symbol $0\left(\Lambda^{-5/2}\right)$ means a term bounded by a constant times $\Lambda^{-5/2}$, for all r.

If we look at the large $\Lambda$ limit, from the point of view of the form factor, we can see that the neighbourhood of the Heisenberg time is critical, in the sense that the following scaling limit exist.

$$\lim_{\Lambda \to \infty} 2\pi\Lambda \left[S(u/\Lambda) - S(u/\infty)\right]\left(u = 2\pi + \frac{2x}{\Lambda}\right) = \frac{1}{\sqrt{\pi}} \int_1^\infty \frac{dt}{t^{3/2}} \exp\left(-t\, x^2\right) \tag{4.14}$$

Let us comment briefly about the comparison of our result, for the large $\Lambda$ limit (Eq. (4.11)) with previous work. We find agreement with the result of Guhr and Müller-Groeling [5] only in the regime $r \ll \Lambda^2$. Our result is consistent with that of perturbation theory [6] if we assume that $r \gg 1$ and average over few level spacing. In this case we get

$$Y_p(r) = -\frac{2\Lambda^2}{r^2 + \left(2\pi\Lambda^2\right)^2} + \frac{1}{2\pi^2 r^2}$$

which corresponds to the first two terms, which are dominant in Eq. (14) of [6] (the $\lambda$ in this reference should be identified with $2\Lambda^2$).

It remains finally to examine the small $\Lambda$ limit. For this purpose, it appears useful to have another integral representation of Y.

Using equations (4.7) and (4.8), one can see that $\frac{d^2 S}{du^2}$ can be expressed in terms of Bessel functions $I_k$. Using recursion formula for these functions and the expression (4.6) for Y, one can express the correlation function in the form

$$Y(r) = -\frac{2\Lambda^2}{r}\int_0^\infty du\ \sin ur\ \frac{2}{\gamma}\left[I_1(\gamma) - \sqrt{\frac{8u}{\pi}}I_2(\gamma)\right]\exp\left[-\Lambda^2 u^2 - 2\pi\Lambda u\right]$$

$$+\frac{2\Lambda^2}{r}\int_0^\infty \frac{\cos ur}{u}\left[I_2(\gamma) - \sqrt{\frac{2u}{\pi}}I_3(\gamma)\right]\exp\left[-\Lambda^2 u^2 - 2\pi\Lambda u\right] \quad (4.15)$$

This expression shows that if one keeps $\frac{r}{\Lambda}$ fixed and let $\Lambda$ tends to zero, one obtains a well defined quantity

$$Y(r) = 1 - \frac{r}{\Lambda}\int_0^\infty du\ \sin u\ \frac{r}{\Lambda}\exp -u^2 + 0\left(\sqrt{\Lambda}\right) \quad (4.16)$$

which coincides with the perturbative result of Leyvraz and Seligman [7].

**Acknowledgements**

One of us (B.S) thanks the EPF-L for hospitality and S. Hikami, M. Janssen, R. Pruni, for useful discussions. This work was supported by a grant from Israel Science Foundation (B.S.) and the "Fonds national suisse de la recherche scientifique" (H. K.).

# Appendix

We give here a short proof of the basic identity 2, used in the text.

Let

$$G(a) = (2\pi\sigma)^{-\frac{N}{2}} \int d\lambda_1 L\, d\lambda_N\, \Delta(\lambda + a)\, \exp\left[-\frac{1}{2\sigma}\sum_{j=1}^{N}\lambda_j^2\right]$$

where

$$\Delta(\lambda + a) = \prod_{N \geq i > j \geq 1}\left(\lambda_i + a_i - \lambda_j - a_j\right)$$

It is easily seen that $G(a)$ is a totally anti symmetric polynomial in the a's of degree N-1 in each variable $a_j$. Therefore

$$G(a) = C_\sigma\, \Delta(a)$$

But since $G(a) = \Delta(a)$ when $\sigma = 0$, we conclude that $G(a) = \Delta(a)$, for all $\sigma$.